\documentclass[11pt]{article}
\usepackage{epsfig}

\title{\bf Production of cumulative pions and percolation of strings}
\author{M.A. Braun\\
 Department of High Energy physics,
 University of S.Petersburg,
198904 S.Petersburg, Russia}
\def\beq{\begin{equation}}
\def\eeq{\end{equation}}

\begin{document}
\maketitle
\medskip
\begin{abstract}
Production of pions in high-energy collisions with nuclei in the kinematics prohibited for free
nucleons ("cumulative pions") is studied in the fusing color string model.The model describes the
so-called direct mechanism for cumulative production. The other, spectator mechanism dominates in production of
cumulative protons but is suppressed for pions. In the model cumulative pions are generated by
string fusion which raises the maximal energy of produced partons above the level of the free nucleon kinematics.
Momentum and multiplicity sum rules are used to determine the spectra in the deep fragmentation region.
Predicted spectra of cumulative pions exponentially fall with the scaling variable $x$ in the interval
$1<x<3$ with a slope of the order 5$\div$5.6, which agrees well with the raw data obtained in the recent
experiment at RHIC with Cu-Au collisioins. However the agreement is worse for the so-called
unfolded data, presumably taking into account corrections due to the expermental set-up and having rather a
power-like form.
\end{abstract}

\section{Introduction}
Production of particles in nuclear collisions
in the kinematical region prohibited in the free nucleon kinematics
("cumulative particles")  has
long aroused interest both from the theoretical and pragmatic point
of views. On the pragmatic side, this phenomenon, in principle, allows to
raise the effective collision energy far beyond the nominal accelerator one.
This may turn out to be very important in the near future, when all
possibilities to construct still more powerful accelerator facilities
become exhausted. Of course one should have in mind that
the production rate
falls very rapidly above the cumulative threshold, so that to use
the cumulative effect for practical purposes high enough luminosity is
necessary. On the theoretical side, the cumulative effect explores the
hadronic matter at high densities, when two or more nucleons overlap
in the nucleus. Such dense clusters may be thought to be in a state
which closely resembles a cold quark-gluon plasma. Thus cumulative
phenomena could serve as an alternative way to produce this new
state of matter.

There  has never been a shortage of models to describe the cumulative
phenomena, from the multiple nucleon scattering mechanism to
repeated hard interquark interactions ~\cite{baldin,strikfurt,efrem,brave1}.
However it should be
acknowledged from the start that the cumulative particle production
is at least in part a soft phenomenon. So it is natural to study it within
the models which explain successfully soft hadronic and nuclear
interactions in the non-cumulative region. Then one could have a
universal description of particle production in all kinematical
regions. The non-cumulative particle production is well described by
the color string models, in which it is assumed that during the
collisions color strings are stretched between the partons of colliding
hadrons (or nuclei), which then decay into more strings and finally
into observed produced hadrons \cite{review}.

As was argued long ago (see e.g.~\cite{brave2} and references therein) that
apart from the slow Fermi motion of nuclear components, absolutely inadequate to explain
the observed cumulative phenomena,
there are basically three mechanisms of the cumulative particle production:
direct, spectator and rescattering. In the direct mechanism
cumulative particles are generated
in the process of collision. In the spectator mechanism (also known as multinucleon correlations inside the nucleus)
cumulative particles exist in the nucleus
by itself, independent of collisions, the role of the latter being just to liberate them. Finally rescattering may move
the initially produced non-cumulitative particle into a cumulative region.
It was found that the role of these three mechanisms is different for different energies and particles.
In particular rescattering can play its role at small energies and degrees of cumulativity but quickly dies out with the growth of both.
The spectator mechanism strongly dominates in the production of cumulative protons. Cumulative pions in contrary are mostly produced
by the direct mechanism.

In the color string picture, for the spectator mechanism to operate, the strings should be formed within the nucleus between its
partons moving at large relative momenta. This is a very different picture as compared with the standard color string approach
when there are no such partons inside the nucleus and string are stretched between partons of the projectile and target.
The common color string picture corresponds to the direct mechanism. So restricting to this picture we hope to describe production
of cumulative pions but not protons produced mostly by the spectator mechanism.

A working model of string fusion and their percolation was proposed by the authors
some
time ago ~\cite{brapaj1,brapaj2}. It proved to be rather successful in explaining a series of
phenomena related to collective effects among the produced strings, such as
damping of the total multiplicity and strange baryon enhancement.
One expects that fusion of strings which enhances the momenta of the produced particles
may also describe production of cumulative particles with momenta far greater than without fusion,
Old preliminary calculations of the production rates
in the cumulative region at comparatively low energies
gave encouraging results  ~\cite{armesto1,armesto,bfmp} . They agree
quite well with the existing data  for production of cumulative pions in hA collisions at
$E_{cm}=27.5$ GeV ~\cite{bayu,niki} but not of cumulative protons for which the cross-section turned out to  far below experiment  However to
pass to higher energies and heavy-ion collisions one has to considerably update
these old treatment .

We stress that the string picture has been introduced initially
to describe particle production in the central region, where the production rate is practically independent of
rapidity but grows with energy. As mentioned, its results
 agree with the data very well \cite{review}.
In contrary cumulative particles are  produced in the fragmentation region. near the kinematical threshold,
where the production rates do depend on rapidity and go to zero at the threshold.
So from the start it is not at all obvious how the color string
approach may give reasonable results also in the deep fragmentation region.
Accordingly an important part of our study is to describe the rate of pion  production
from the initial and fused strings valid in the fragmentation region.
To this aim we shall use color and momentum conservation imposed on the average
and sum rules which follow.

 As we shall see from our results,
we reproduce a very reasonable description
of the pion production rates  for  $1<x<2$ at 27.5 GeV [5]. However we are not
able to describe the proton rates, which remain experimentally
two orders of magnitude greater than our predictions. As explained  the bulk
of cumulative protons come from the spectator  mechanism, which lies outside our
color string dynamics.
Note that the spectator mechanism  was included in the Monte-Carlo code of ~\cite{armesto} where
 it gave results for nucleon production in agreement
with the experimental data.

The bulk of our paper is devoted to production of cumulative pions at in AA collisions
at RHIC and LHC facilities, related to the performed and planned experimental efforts in this direction.
It is to be noted that in older calculational models HIJING ~\cite{hijing} and DPMJET ~\cite{dpmjet},
devoted to the overall spectra in heavy-ion collisions, particles emitted with energies up to 2$\div$2.5 times
greater than allowed
by proton-proton kinematics were found.
A recent experimental study devoted specifically to cumulative jet production was performed for Cu-Au collisions at 200 Gev
~\cite{bland}. Comparison of our predictions with these data will be postponed until the  discussion
at the end of our paper. With certain reservations the data
confirm the universality of particle production in the fragmentation
region and in particular in the cumulative region.

\section{The model}
The color string model assumes that each of the colliding
hadrons consists of partons (valence and sea quarks), distributed both
in rapidity and transverse space with a certain probability, deduced
from the experimentally known transverse structure and certain theoretical
information as to the behavior of the $x$ distributions at its ends.
These distributions are taken to be the ones for the endpoints
of the generated strings. As a result, the strings acquire a certain length
in rapidity. We shall choose the c.m. system for the colliding nucleons
with the nucleus (projectile) consisting of $A$ nucleons and moving in the forward direction.
Each of the projectile nucleons is taken to carry momentum $p_1$, so that the total momentum
of the projectile nucleus is $Ap_1$. The target is assumed to be just the nucleon with momentum $p_2$.
The cumulative
particles will be observed in the forward hemisphere in the $z$ direction of the fast moving nucleus.
Their longitudinal "+" momenta will be $x_+p_{1=}$ with $x_+>1$. In the following $x_+$ will be called cumulatiity index,
or simply cumulativity. Theoretically the maximal value for $x_+$ is $A$ but in practice we find $x_+\leq 5$.

The nucleons for both projectile and target are split into partons as shown in Figs. 1 and \ref{cumab} for the projectile
where the partons (quarks and diquarks) are illustrated by dashed lines. Color strings are stretched between partons of
the projectile and targets as in Fig. 1 and some of these simple strings can be fused into strings with more color.
In Fig. \ref{cumab} it is shown that the initial 4 simple strings combine  into  fused strings attached to  quark-antiquark pairs
within  the same nucleons (left)  or different nucleons (right) in the projectile nucleus.
\begin{figure}
\begin{center}
\epsfig{file=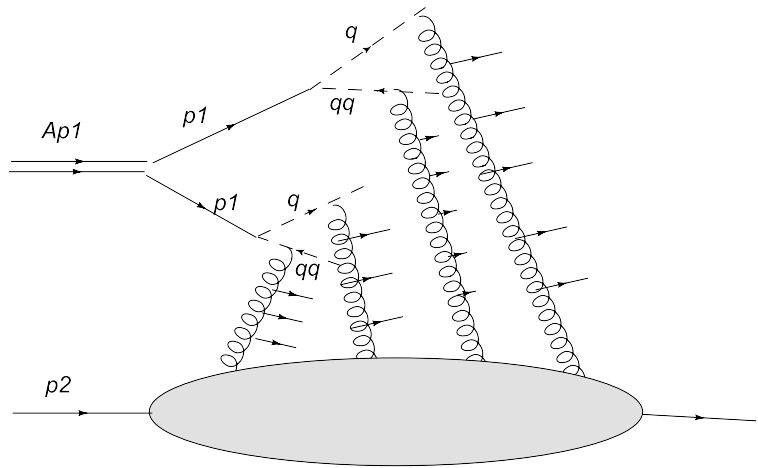, width=10 cm}
\caption{$pA$ collision for $A=2$ with creation of 4 color strings. Nucleons of the
projectile are shown by solid lines,
partons in which they split (quarks and diquarks) by dashed lines}
\end{center}
\label{cum1}
\end{figure}

\begin{figure}
\begin{center}
\epsfig{file=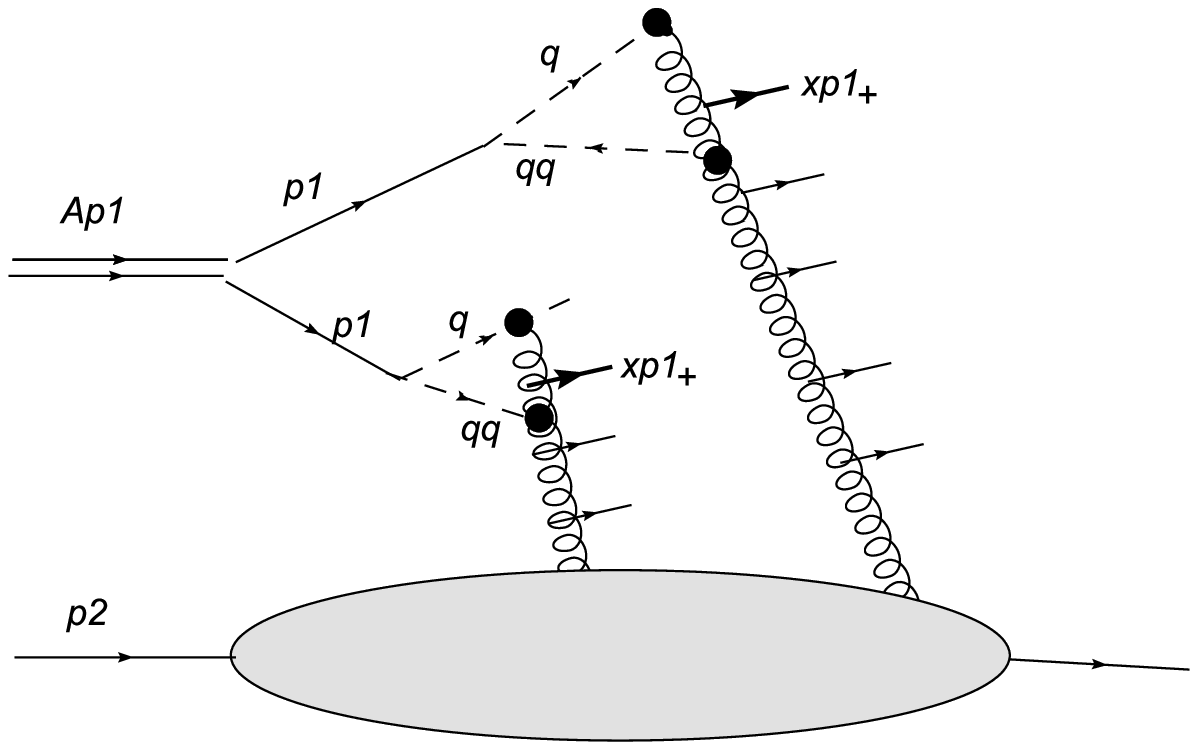, width=7 cm}
\epsfig{file=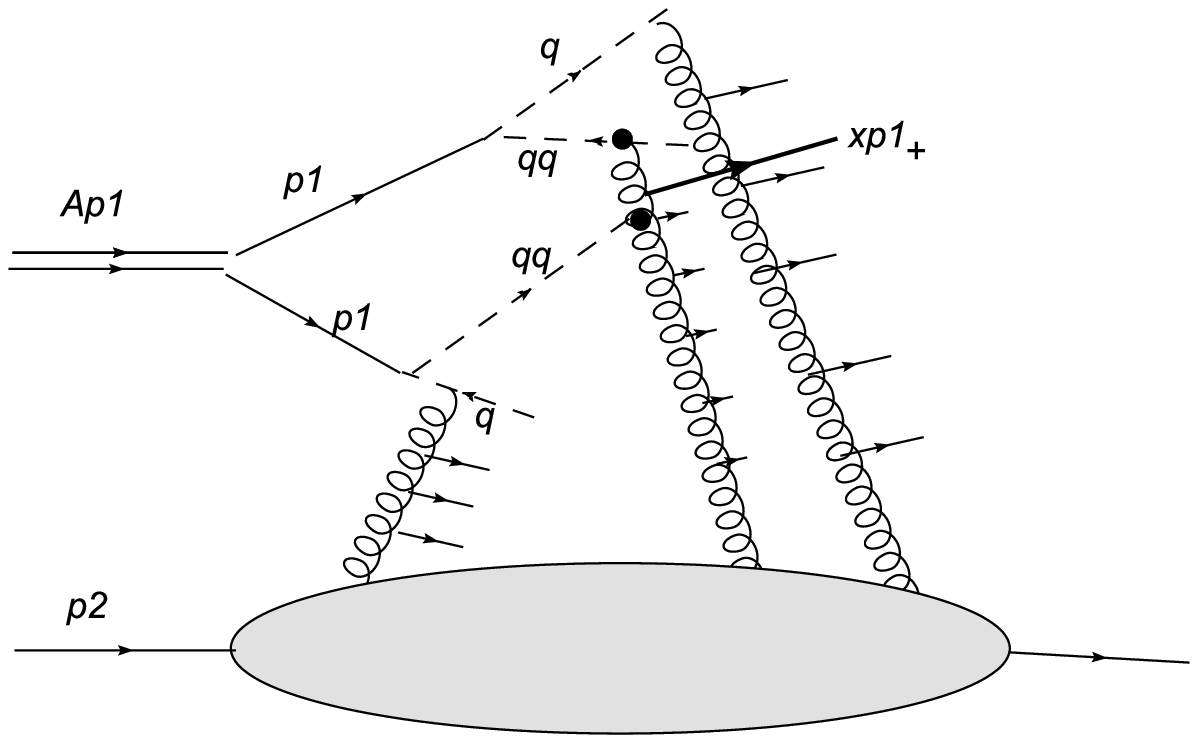, width=7 cm}
\caption{$pA$ collision for $A=2$ with creation of 4 color strings which fuse within individual nucleons (left panel) or
between different nucleons (right panel). Nucleons of the
projectile are shown by solid lines,
partons in which they split (quarks and diquarks) by dashed lines. Cumulative particles are shown by thick solid lines.}
\end{center}
\label{cumab}
\end{figure}

Let a parton from the projectile carry a part
$x_{1+}$ of the "+" component of nucleon momentum $p_1$ and
a partner parton from the target carry a part $x_{2-}$ of the "-" component of
nucleon momentum $p_2$. The total energy squared for the colliding
pair of nucleons is
\beq S=2p_{1+}p_{2-}=m^2e^Y\eeq
where $m$ is the nucleon mass
 and $Y$ is the total rapidity. (We assume that the energy is high, so tat $S>>m^2$).
The c.m. energy squared  accumulated in the string is then
\beq s= x_{1+} x_{2-}S.\eeq
Note that the concept of a string has only sense in the case when $s$
 is not too small, say more than $m^2$. So both $ x_{1+}$ and $x_{2-}$ cannot be
too small.
\beq  x_{1+}, x_{2-}>x_{min}=m/\sqrt{S}=e^{-Y/2},\eeq
We relate the scaling variables for the string endpoints
to their rapidities by
\beq
y_1=Y/2+\ln x_{1+},\ \ y_2=-Y/2-\ln x_{2-}.
\label{y12}
\eeq
Due to (\ref{y12}) $y_1\geq 0$ and $y_2\leq 0$. The "length" of the string is just
the difference $y_1-y_2$.

Due to  partonic distribution in $x$ the strings have
different lengths and moreover can take different position in rapidity respective to
the center $y=0$.
The sea distribution in a hadron is much softer than the valence one.
In fact the sea distribution behaves as $1/x$ near $x=0$, so that the
average value of $x$ for sea partons is small, of the order $x_{min}$
~\cite{capella}.
As a result, strings attached to sea partons in the projectile nucleus
carry very small parts of longitudinal momentum in the forward direction,
which
moreover fall with energy, so that they seem to be useless for building
up the cumulative particles.
This allows us to  retain only strings
attached to valence partons, quarks and diquarks, in the projectile and
neglect all  strings attached to sea quarks altogether. This is reflected in Fig. \ref{cum1}
where we have shown only the valence partons in the projectile.
Note that the number of the former is exactly equal to $2A$ and does not
change with energy. So for a given nucleus we shall have a fixed number of
strings, independent of the energy.

The upper end rapidities of the
strings attached to diquarks are usually thought to
be  larger than of those attached to the quarks, since the average
value of $x$ for the diquark is substantially larger  that for the
quark. Theoretical considerations lead to the conclusion  that as
$x\to 1$ the distributions for the quark and diquark in the nucleon
behave as $(1-x)^{3/2}$  and $(1-x)^{-1/2}$ respectively, modulo
logarithms ~\cite{capella}. Neglecting the logarithms and taking also in account the
behavior at $x\to 0$ we
assume that these distributions for the quark$q$ and diquark$qq$ are
\beq
q(x)=\frac{8}{3\pi}x^{-1/2}(1-x)^{3/2}
\label{q}
\eeq and
\beq
qq(x)=q(1-x)=\frac{8}{3\pi}x^{3/2}(1-x)^{-1/2}
\label{qq}
\eeq

The quark and diquark strings will be attached  to all sorts of partons
in the target nucleon: valence
quark and diquark and sea quarks. Their position in rapidity in the backward
hemisphere will be very different.  However we are
 not interested in the spectrum in the backward hemisphere. So, for our
purpose, limiting ourselves with the forward hemisphere, we
may take lower  ends of the strings all equal to $x_{min}<<1$.
As a result, in our  model at the start we have $2A$ initially created
strings, half of them attached to quarks and  half to diquarks, their
lower ends in rapidity all equal to \[y_2=Y/2+\ln x_{min}\]
and their upper ends distributed in accordance with (5) and (6).
As soon as they overlap in the transverse space they fuse into new
strings with more color and more energy. This process will be studied in
the next section.

\section{Fragmentation spectra and fusion of strings}
\subsection{One sort of strings and particles}
The following discussion closely follows that of ~\cite{bfmp}.
To start we shall study a simplified situation with only one sort of strings.
We shall consider both the original string stretched between the partons of the projectile and target
and the fused strings of higher color which are generated when $n$ original strings occupy the same area in the
transverse plane.

Consider first the original simple string.
Let it  have its ends at $x_{1+}\equiv x_1$ and $x_{min}$. For cumulative particles
 we shall be interested only in the forward hemisphere and only in "+" components of momenta
so that in the following we omit subindex "+".

We shall be interested in the spectrum of particles emitted from
this string with the longitudinal momentum $xp_1$. Evidently  $x$ varies in the interval
\[ x_{min}<x<x_1,\]
or introducing
$z=x/x_1$ in the interval
\[z_{min}<z<1,\ \ z_{min}=\frac{x_{min}}{x_1}.\]
The multiplicity density of produced particles (pions) will be
\[
\tau_1(z)=\frac{d\mu}{dy}
\]
and the total multiplicity of particles emitted in one of the two hemispheres
is
\[\int_{z_{min}}^1dz\tau_1(z)=\frac{1}{2}\mu_0,
\]
where $\mu_0$ is the total multiplicity in both hemispheres.
The emitted particles will have their "+" momenta $k_+$ in the interval
\[x_{min}p_{1+}<k_+<zx_1p_{1+}\]
and since $z,x_1\leq 1$
\[x_{min}p_{1+}<k_+<p_{1+}.\]
So the particles emitted from the simple string cannot carry their "+" momenta greater than a single incoming nucleon.
They are non-cumulative.

Now let several simple strings coexist without fusion. Each of these strings will produce particles in the interval
dictated by its ends. If the $i$-th string has its upper end $x_1^{(i)}$ then the total multiplicity density of $n$
not fused strings will be
\[\tau^{(n)}(x)=\sum_{i=1}^n\tau_1^{(i)}(x),\]
where $\tau_1^{(i)}$ is the multiplicity density of the $i$-th string, different from zero in the interval
\[x_{min}<x<x_1^{(i)}\leq 1.\]
As  a result all produced particles will have their "+" momenta lying in the same interval $<p_+$ as for a single string,
so that they all will be non-cumulative. We conclude that there will not appear any cumulative particles without string fusion.
Only fusion of strings produces cumulative particles in our picture.

Now consider that $n$ simple strings fuse into a fused string.
 The process of fusion obeys two conservation
laws: those of color and momentum.
As a result of the conservation of color, the color of the fused string is
$\sqrt{n}$ higher than that of the ordinary string [7,8]. From the 4 momentum
conservation laws we shall be interested mostly in the conservation of the
"+" component, which leads to the conservation of $x$.  The fused string
will have its upper endpoint
 \[x_n=\sum_{i=1}^nx^{(i)},\]
where $x^{(i)}$ are upper ends of
fusing strings.
This endpoint can be much higher than individual $x^{(i)}$ of the fusing strings.
In the limiting case when each  fusing string has $x^{(i)}=1$ we find
$x_n=n$. Consequently the particles emitted from the fused string will have their
maximal "+" momentum $np_+$  and be cumulative with the degree $n$ of cumulativity.

At this point we have to stress that there are some notable exceptions. The maximal value
$n$ for $x_n$ can be achieved only when  different strings which fuse are truly independent,
which is so if the strings belong to different nucleons in the projectile. To see it
imagine that two string fuse which belong to the same nucleon (one starting from the
quark and the other from antiquark). In this case $x^{(1)}+x^{(2)}=1$ and $x_2$ will have the same value as $x_1$.
So fusing of strings inside the nucleon does not give any cumulative particles. Such particles are only generated
by fusing of strings belonging to different nucleons in the projectile, compare let and right panels in Fig \ref{cumab}.

The multiplicity density of particles emitted from the fused string will be denoted
 \[ \tau_n=\frac{d\mu_n}{dy}\]
where $\mu$ is the generated multiplicity.
It is different from zero in the interval
\beq
x_{min}^{(n)}\leq x\leq x_n,\ \ x_{min}^{(n)}=nx_{\min},\label{intx}
\eeq
or again introducing
$z=x/x_n$
in the interval
\[z_{min}\leq z<1,\ \ z_{min}=\frac{x^{(n)}_{min}}{x_n}\]
We are interested in emission at high values of $x$, or $z$ close to unity,
that is in the fragmentation region for the projectile. Standardly it is assumed that the multiplicity density is
practically independent of $x$ in the central region, that is at small $x$.
However $\tau_n$ cannot be constant in the whole interval (\ref{intx}) but has to approach zero
at its end in the fragmentation regions.
At such values  of $z$
$\tau_n$ is expected to strongly depend on $z$,
Our task is to formulate the $z$ dependence of $\tau_n$ in this kinematical region.

To this aim we set up  certain sum rules which follow from the mentioned conservation laws and
restrict possible forms of the spectrum of produced hadrons.

The total
number of particles produced in the forward hemisphere by the fused string
should be $\sqrt{n}$ greater than by the ordinary string. This leads to
the multiplicity sum rule:
\beq
\int_{nx_{min}}^{x_n}\frac{dx}{x}\tau_n(x)=\frac{1}{2}\mu_0\sqrt{n}
\eeq
where as before $\mu_0$ the total multiplicity from a simple string in both
hemispheres.
The produced particles have to carry all the longitudinal momentum in the
forward direction. This results  in the  sum rule for $x$:
\beq
\int_{nx_{min}}^{x_n}dx\tau_n(x)=x_n
\eeq
In these sum rules $x_{min}$ is given by (3) and is small.
Passing to the scaled variable
\[z=x/x_n\]
 we rewrite the two sum rules
as
\beq
\int_{z_n}^1\frac{dz}{z}\tau_n(z)=\frac{1}{2}\mu_0\sqrt{n}
\eeq
and
\beq
\int_{z_n}^1dz\tau_n(z)=1
\eeq
where
\beq
z_n=nx_{min}/x_n
\eeq

These sum rules put severe restrictions on the form of the distribution
$\tau_n$, which obviously cannot be independent of $n$. Comparing (7) and
(8) we see that the spectrum of the fused string has to vanish at its
upper threshold faster than for the
simple string. In the scaled variable $z$ it is shifted to smaller values
(and thus to the central region).
This must have a negative effect on the formation of cumulative particles
produced at the extreme values of $x$.

To proceed, we choose the simplest form for the distribution $\tau_n$:
\beq
\tau_n(z)=a_n(1-z)^{\alpha_n-1},\ \ \alpha>1
\label{tn}
\eeq
with only two parameters magnitude $a_n$ and slope $\alpha_n$.

The $x$ sum rule relates $a_n$ and $\alpha_n$:
 \beq
 a_n=\alpha_n(1-z_n)^{-\alpha_n}.
 \label{an}
\eeq
The multiplicity sum rule finally determines $\alpha_n$  via $\mu_0$:
\beq
\alpha_n(1-z_n)^{-\alpha_n}\int_{z_n}^1\frac{dz}{z}(1-z)^{\alpha_n-1}=
\frac{1}{2}\mu_0\sqrt{n}
\eeq

This equation can be easily solved when $z_n\to 0$
We present the integral in (14) as
\beq
\int_{z_n}^1\frac{dz}{z}[(1-z)^{\alpha_n-1}-1]+\ln\frac{1}{z_n}.
\eeq
The integral term is finite at $z_n=0$ so that we can write it as a
difference
of integrals in the intervals $[0,1]$ and $[0,z_n]$. The first can be found
exactly
\[
I_1=\int_0^1\frac{dz}{z}[(1-z)^{\alpha_n-1}-1]=
\lim_{\epsilon\to 0}
\int_0^1dzz^{-1+\epsilon}[(1-z)^{\alpha_n-1}-1]=\]\beq
\lim_{\epsilon\to 0}\Big[{\rm B}(\alpha_n,\epsilon)-\frac{1}{\epsilon}\Big]=
\psi(1)-\psi(\alpha_n).
\eeq
The second term has an order
$-(\alpha_n-1)z_n$
and is small unless $\alpha_n$ grows faster than $n$, which is not the case
as we shall presently see. In fact we shall find that $\alpha_n$ grows roughly
as $\sqrt{n}$, which allows to neglect the second factor in (14) and rewrite
it in its final form
\beq
\alpha_n \Big[\ln\frac{1}{z_n}+\psi(1)-\psi(\alpha_n)\Big]=
\frac{1}{2}\mu_0\sqrt{n}.
\label{aln}
\eeq
Note that the total multiplicity $\mu_0$ from a simple string is just $Y$.
Also $1/z_n=nx_{min}/x_n$ and so
\[\ln \frac{1}{z_n}=\frac{Y}{2}+\ln\frac{x_n}{n}\]
So Eq. (\ref{aln}) can be rewritten as
\beq
\alpha_n=\sqrt{n}\Big(1+\frac{2}{Y}(\ln\frac{x_n}{n}+\psi(1)-\psi(\alpha_n))\Big)^{-1}
\label{aln1}
\eeq

This transcendental equation determines $\alpha_n(x_n)$ for the fused string.
Obviously at $Y>>1$ the solution does not depend on $x_n$ and is just $\alpha_n=\sqrt{n}$.
 To finally fix the distributions at finite $Y$ we have to
choose the value of
$\alpha$ for the simple string. We take the simplest choice $\alpha_1=1$
for an average string with $x=x_0=1/2$,
which corresponds to a completely flat spectrum and agrees with the
results of ~\cite{capella}.  This fixes the multiplicity density for the average string
\beq\tau_1(y)=1\eeq
which favorably compares to the value 1.1 extracted from the experimental data
[8]. After that the equation for $\alpha$ takes the form
\beq
\alpha_n\Big(\ln x_n+\psi(1)-\psi(\alpha_n)\Big)=\sqrt{n}
\label{aln2}
\eeq
At finite $Y$ it has to be solved numerically to give $\alpha_n(x_n,Y)$
where $x_n$ is the upper end of the string $n$

We  find that with the growing $n$ the spectrum of produced
particles
goes to zero at $z\to 1$ more and more rapidly. So although strings with large
$n$ produce particles with large values of $x\le x_n$, the production rate
is increasingly small.

\subsection{Different strings and particles}

In reality strings are of two different types, attached to quarks or antiquarks.
Also various types of hadrons are produced in general. In the cumulative
region the mostly studied particles are nucleons and pions, the production
rates of the rest being much smaller. As mentioned in the Introduction the
dominant mechanism for emission of cumulative nucleons is the spectator one, which
lies outside the color string picture. So we restrict ourselves to cumulative pions.
The multiplicity densities for each sort of fused string will obviously
depend on its flavor contents, that is, of the
number of quark and diquark strings in it.

Let the string be composed of $n-k$ quarks and $k$ diquarks, $k=0,1,...n$
 We shall then have distributions
$\tau_{nk}$ for the produced pions.
The multiplicity and momentum sum rules alone are now insufficient to
determine each of the distribution $\tau_{nk}$ separately.
To overcome this difficulty we note that  in our picture the observed
pion is produced when the  parton (quark or diquark) emerging from
string decay neutralizes  its color by picking up an appropriate parton
from the vacuum. In this way a quark may go into a pion if it picks up
an antiquark or into a nucleon if it picks up two quarks. The quark
counting rules
tell  us that the behavior at the threshold in the second case will
have two extra powers of $(x-x_n)$. Likewise a diquark may go either
into a nucleon picking up a third quark or into two pions picking up
two antiquarks, with a probability smaller by a factor $(x-x_n)^2$ at
the threshold.  On the other hand at the threshold the probability to find
a quark in the proton is $(1-x)^2$ smaller than that of the diquark , Eqs.
(\ref{q}) and (\ref{qq}).  The two effects, that of color neutralization and
threshold damping in the nucleus,  seem to  compensate each other.
So that in the end  the
pion production rate from the  antiquark string is  just twice the rate from the quark string
provided the distribution of the former in the nucleus is the same as for the quark strings.
This enables us to take the same distributions (\ref{q}) for quark and antiquark strings in the nucleus
and for the fragmentation function $\tau_{nk}$ use
\beq
\tau_{nk}=\tau_n\Big(1+\frac{k}{n}\Big)
\label{tnk}
\eeq
where $\tau_n$ is the distribution (\ref{tn}) determined in the previous subsection.
Eq. (\ref{tnk}) takes into account  doubling of the  pion production from antiquark strings.
For the simple string it correctly gives
\[\tau_{10}=\tau_1,\ \ \tau_{11}=2\tau_1\]

Averaging (\ref{tnk}) over all $n$-fold fused strings one has the average $<k>=n/2$ , so that
$\tau_{nk}$ can be well approximated by
\beq
\tau_{nk}=\frac{3}{2} \tau_n
\label{tnk1}
\eeq

Note that should we want to consider cumulative protons then  quark strings would give practically no contribution being
damped both at the moments of their formation and neutralization of color. The antiquarks in contrast will
dominate at both steps and give practically the total contribution. So then one has to consider only antiquark strings and only
one multiplicity distribution, that of nucleons $\tau_n^{(N)}$ for which our sum rules will be valid with the only change
$\mu_0\to\mu_o{(N)}$, the total multiplicity of nucleons. However one then have to use distribution (\ref{qq})
for antiquark strings in the nucleus which grows in the fragmentation region.

\section{Nucleus-nucleus scattering}
In the preceding sections we studied $pA$ scattering in the system where the nucleus is moving fast in the positive direction $z$.
Correspondingly we were  interested in the forward  hemisphere in the deep fragmentation region with attention to particles emitted with
longitudinal momenta higher than that of the projectile nucleons. The role of the target proton was purely spectatorial, since
we were not interested in particles moving in the opposite direction to the projectile nucleus. The only information necessary about the target
was that all  strings attached to the projectile nucleus could  be attached to the target. This was related to existence of sea partons in
the target apart from the dominant valence ones.

If  one substitutes the proton target by the nucleus (say of the same atomic number $A$) nothing will change in the
projectile nucleus hemisphere, so that all our previous formulas remain valid. The only difference will be that  strings attached to the nucleons
in the projectile nucleus can now be coupled to  valence partons in the target nucleus provided both nuclei overlap in the transverse area.
So the number
of all strings will depend on geometry, more concretely on the impact parametr $b$. As for pA collisions, formation of cumulative strings with $x>1$
will require that fusing strings belong to different nucleons in the projectile, So the picture of cumulative production will not change,
except that it will be different for
different $b$. The final cumulative multiplicity will be obtained as usual by integration over all values of impact parameter $b$.

Thus so far as the cumulative particles are concerned the difference between $pA$ and $AA$ collisions reduces to
the geometry in the transverse plane and the ensuing change of string configurations.

\section{Probability of cumulative strings}
\subsection{Geometric probability of string fusion}
As stressed, the cumulative production in our scenario is totally explained by  formation
of fused strings, which follows when $n\geq 2$ strings overlap in the transverse space.
The exact nature of this overlapping may be different, total or partial. In the transverse space
such fused strings may have different forms and dimensions thus presenting complicated geometrical structures.
The detailed analysis of their geometry and dynamical properties presents an exceptionally complicated and hardly realizable task
even when the number of strings is quite small, to say nothing of the realistic case when this number
is counted by hundreds or even thousands. However the study of cases with a small number of strings
shows  that equivalent results can be well reproduced within a simplified picture ~\cite{bkpv}.
Cover the transverse area of interaction by a lattice with cells having areas of the simple strings
(circles of the radius $\sim 0.3$ fm). Strings stretched between the projectile and target turn out to appear
 in one or different cells. Once some cell contains $n$ strings then they are assumed to fuse and give rise to
 a $n$-fold fused string occupying this cell.

 In this approach formation of fused strings proceeds in several steps.  Consider pA collisions.
 At the first step one sets up the mentioned  lattice to cover all nucleus area.
Cells form the file $z_c(m)$ $m=1,2,,$ of their points $z_c=(x,y)=x+iy$ in the transverse plane with the center of the
nucleus at $z_A=(0,0)$.

Second step is to throw randomly $A$ nucleons at points $z_N$ with the probability given by the
transverse density $T(b)$.
They are thrown successively and with each new nucleon one passes to

Third step which  is to randomly throw 2 strings around each of the thrown nucleon
at distance from its center dictated by the appropriate matter density within the nucleon (Gaussian).
Each of the two strings  then arrives into some cell $m$ which enhances its string content
$\nu(m)$, $\nu=0,1,2,..$ by unity.

At this point one has to take into account that two fusing strings, quark and antiquark,
attached to the same nucleon in the projectile nucleus do not generate a cumulative string with their
upper end $x_n>1$ (see Section 3.1).
They are to be excluded from the total set of fused strings leaving only those which are generated
when to the target  two strings from different nucleons of the projectile nucleus are attached.
To do this we note that
the two strings from the same nucleon may be put in different cells $m_1$ and $m_2$ or
in the same cell when $m_1=m_2$. In the former case both $\nu(m_1$ and $\nu(m_2$ are
each enhanced by unity. In the latter case $\nu(m_1)=\nu(m_2)$ does not change.
As a result in the cumulative production a fused string is only  generated when it belong to different nucleons in the
projectile nucleus, so that they have to overlap in the transverse area. This introduces factor of smallness, roughly the ratio of the
transverse areas of the nucleon to nucleus for each successive fusion of $n=2,3,...$ strings and is responsible for the fast
decrease of the cumulative cross-section with the growth of cumulatively number $x$.

At the fourth step one searches all cells with more than 2 strings. One finds $N_c(2)$  cells with 2 strings,
that is 2-fold fused strings,
$N_c(3)$ cells with 3 strings, that is 3-fold fused strings, and so on.
Different cells mark overlap of several nucleons at different locations in
the transverse plane and physically to different trajectories of the target proton at each collision.
So to find the total cross-section on has to take the sum of  contributions of all cells with a given number of $n$  strings calculated with the
relevant dynamical probability $p_n$ and particle distribution $\tau_n(z)$ Eq,(\ref{tn}). This corresponds to the cross-section at all impact
parameters $b$
of the  target proton as it crosses the nucleus.

One has to recall that the cumulative string with $x_n>1$
can only be formed when it starts from valence quarks in the projectile nucleus. So only two strings can be attached to
each nucleon and the total number of simple strings is fixed to $2A$. In the contrary for non-cumulative production
also sea quarks in the projectile contribute with the number od strings from each nucleon and the resulting multiplicity
steadily increasing with energy. From this one immediately concludes that cumulative production depends only very weakly on energy,
all dependence coming from powers $\alpha_n(x,Y)$.

For AA scattering the procedure does not change with the only difference that the projectile nucleus is substituted by the nuclei overlap
depending on the impact parameter $b$ for the collision. So one will get different cumulative multiplicity for different $b$.
The total multiplicity will be obtained after integrations over all $b$.

\subsection{Probability of cumulativity of the fused string}

Strings are distributed in the nucleons with probabilities (\ref{q}) and (\ref{qq}) for quarks and diquarks.
As argued above we shall assume that they all be distributed with the quark distribution (\ref{q}). To eliminate the
steep growth at $x=0$ we pass to variable $u=\sqrt{x}$. In terms of $u$ the distribution takes the simple form
\beq
\rho(u)=(1-u^2)^{3/2}=(1-x)^{3/2},\ \ 0<u<1.
\label{rho}
\eeq
The probability to find a string with the upper end at $x$ consisted of $n$ simple ones with ends $x_1,...x_n$ is
given by the multiple integral
\beq
p_n(x)=\int_0^1\prod_{i=1}^n\Big(du_i\rho(x_i)\Big)\delta\Big(x-\sum_{i=1}^n x_i\Big)
=\int_0^1\prod_{i=1}^{n-1}\Big(du_i\rho(x_i)\Big)\rho(\Big(x-\sum_{i=1}^{n-1}x_i\Big).
\label{pn}
\eeq
We have  to determine the limits of successive integrations
starting from the $(n-1)$-th. From the start it is obvious that $p_n(x)$ can be different from zero only
in the interval $0<x<n$.

For two strings we have
\beq
p_2(x)=\int_0^1du_1\rho(x_1)\rho(x-x_1),\ \ u_1=\sqrt{x_1}.
\label{p2}
\eeq
Evidently we should have
\[0<x_2=x-x_1<1,\ \ {\rm or}\ \ 0<x-x_1,\ \ x-x_1<1.\]
These two conditions determine the lower and upper limits $a_1$ and $b_1$ of integration over $u_1$:
\beq
x_1>a_1(x)=\max(x-1,0),\ \ x_1<b_1(x)=\min(x, 1),
\label{ab2}
\eeq
or correspondingly the limits in $u_1$
\beq
\sqrt{a_1(x)}<u_1<\sqrt{b_1(x)}.
\label{ab2u}
\eeq
Probability $p_2(x)$ is different from zero in the region of $x$ such that
$a_1(x)<b_1(x)$.\\
If $0<x<1$ then $a_1=0$ and $b_1=x$ so $a_1<b_1$,\\
if $1<x<2$ then $a_1=x-1$ and $b_1=1$ so $a_1<b_1$ provided $x<2$.\\
If $x>2$ then $a_1=x-1$ and $b_1=1$ so $a_1>b_1$ and $p_2(x)=0$, as noted previously.\\
As a result a nonzero result is obtained at $0<x<2$ but with different integration
limits. In the case of interest $x>1$ the limits are $a_1=x-1$ and $b_1=1$.

Now consider $p_n(x)$ for $n>2$\\
From (\ref{pn}) one finds a recurrent relation
\beq
p_n(x)=\int_{u_{min}}^{u_{max}}du_1\rho(x_1)p_{n-1}(x-x_1).,\ \ u_1=\sqrt{x_1}.
\label{pnr}
\eeq
The limits
\[u_{min}=\sqrt{a}\ \ u_{max}=\sqrt{b}\]
are determined by the condition $p_{n-1}(x-x_1)\neq 0$, which limits $x-x_1$ to the region
\[0<x-x_1<n-1.\]
From this we find
\[x_1<b=\min(x,1),\ \ x_1>a=\max(x-n+1,0).\]
For $x>1$ it follows that $b=1$ is independent of $x$\\
As to $a$ for $x<n-1$ we get $a=0$. However in the interval $n-1<x<n$ we obtain $a=x-n+1$.

Formula (\ref{pnr}) can be used for the  calculation of $p_n(x)$  starting from $p_2(x)$
explicitly given by integral (\ref{p2}).

\section{Calculations}
For both proton-nucleus and nucleus-nucleus collisions one has to know
the probabilities of fused strings formation $p_n(x)$ and observed particle distribution $\tau_n(x)$. The former are determined by
Eq. (\ref{p2}) and recurrent relation (\ref{pnr}).
The latter are fully expressed by the
powers $\alpha_n(x,Y)$, which in turn are determined by Eq. (\ref{aln2}).
In fact  fused strings formed from more than 5 simple strings are not found in our calculations both for p-A and A-A collisions.
For $n=1,2,..5$ the results of our numerical calculations give $p_n(x)$ and $\alpha_n(x,Y)$ which are presented in
Figs. 3,4 and 5

\begin{figure}
\label{fig2}
\begin{center}
\epsfig{file=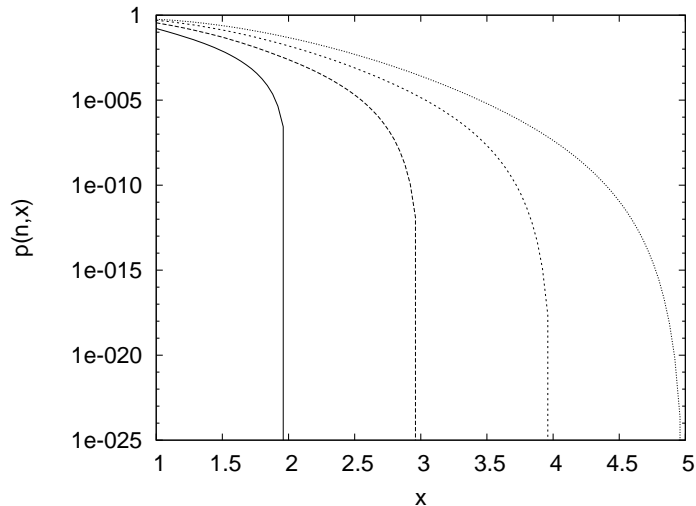, width=10 cm}
\caption{Probabilities $p_n(x)$. They are different from zero in the interval at $1<x<n$}
\end{center}
\end{figure}

\begin{figure}
\begin{center}
\epsfig{file=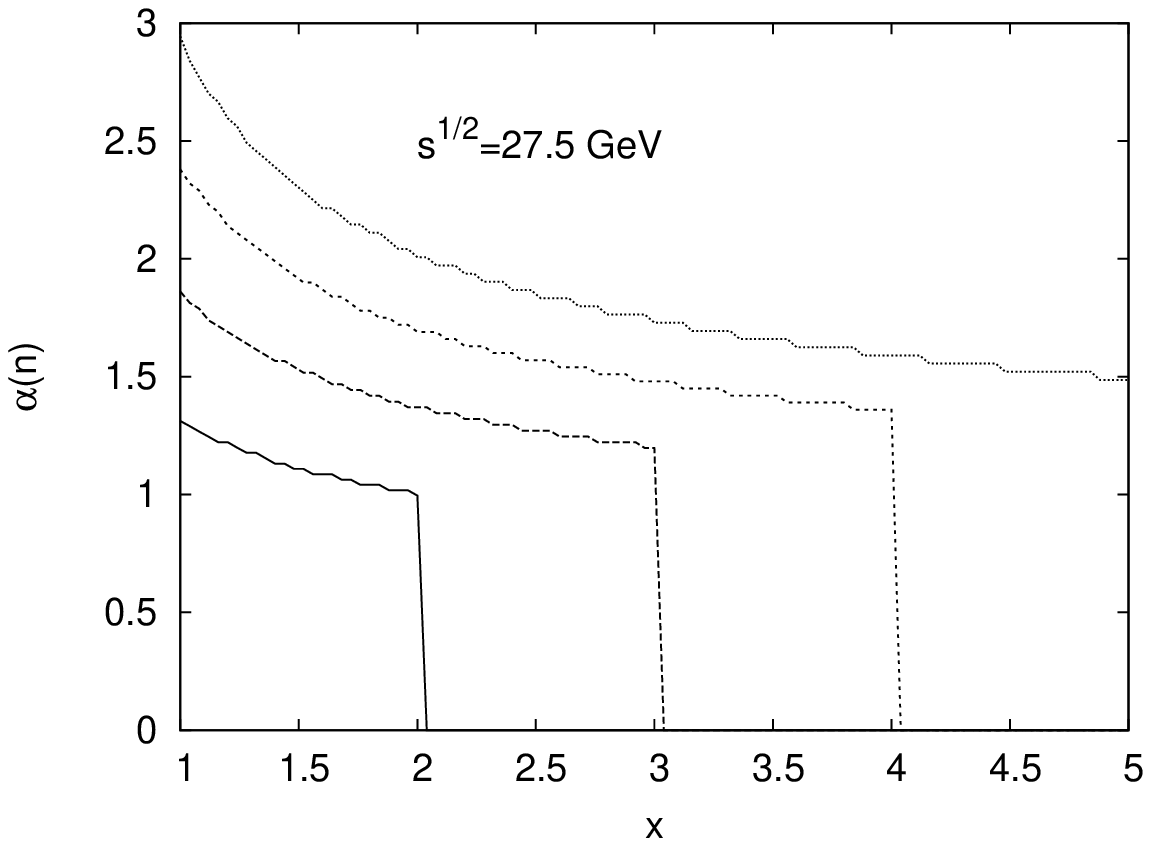, width=7 cm}
\epsfig{file=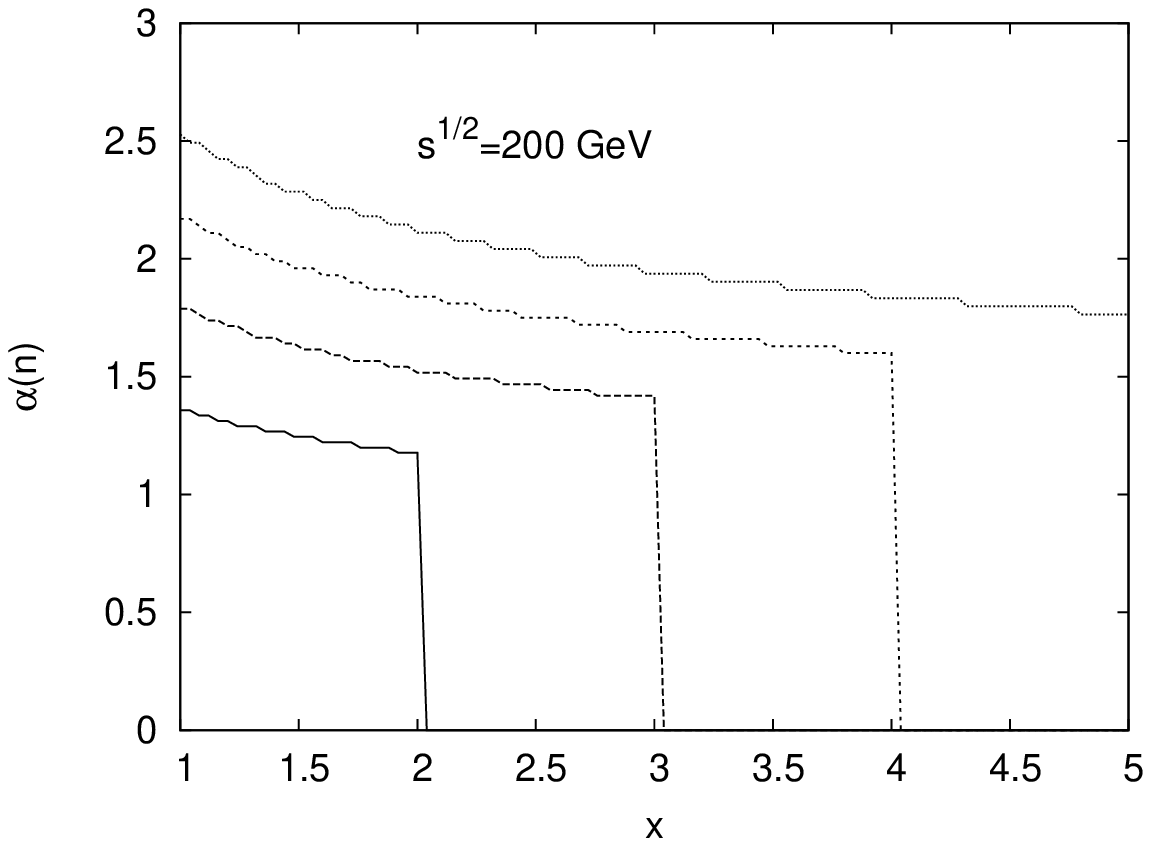, width=7 cm}
\caption{Powers $\alpha_n(x)$ at $\sqrt{s}=$ 27.5 and 200 GeV.
Curves from bottom to top correspond to $n=2,3,4,5$}
\end{center}
\label{fig3}
\end{figure}

\begin{figure}
\begin{center}
\epsfig{file=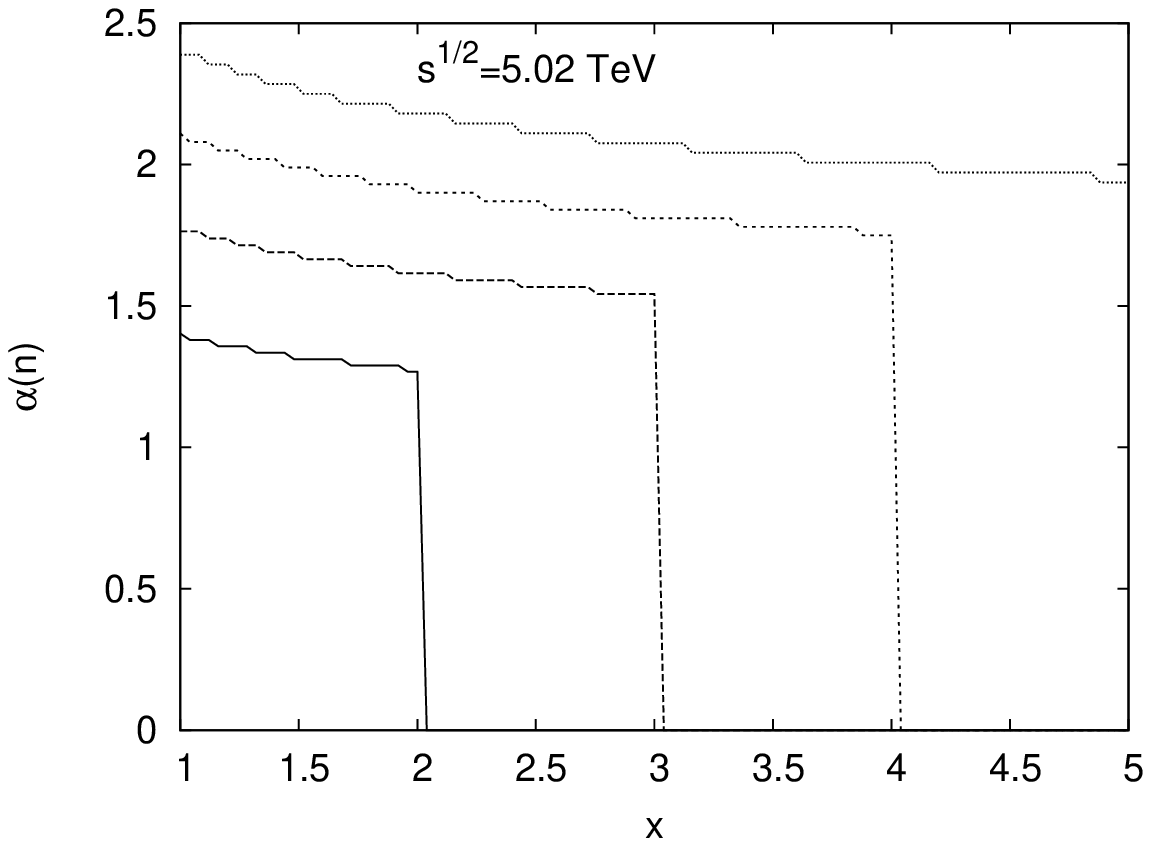,width=7 cm}
\epsfig{file=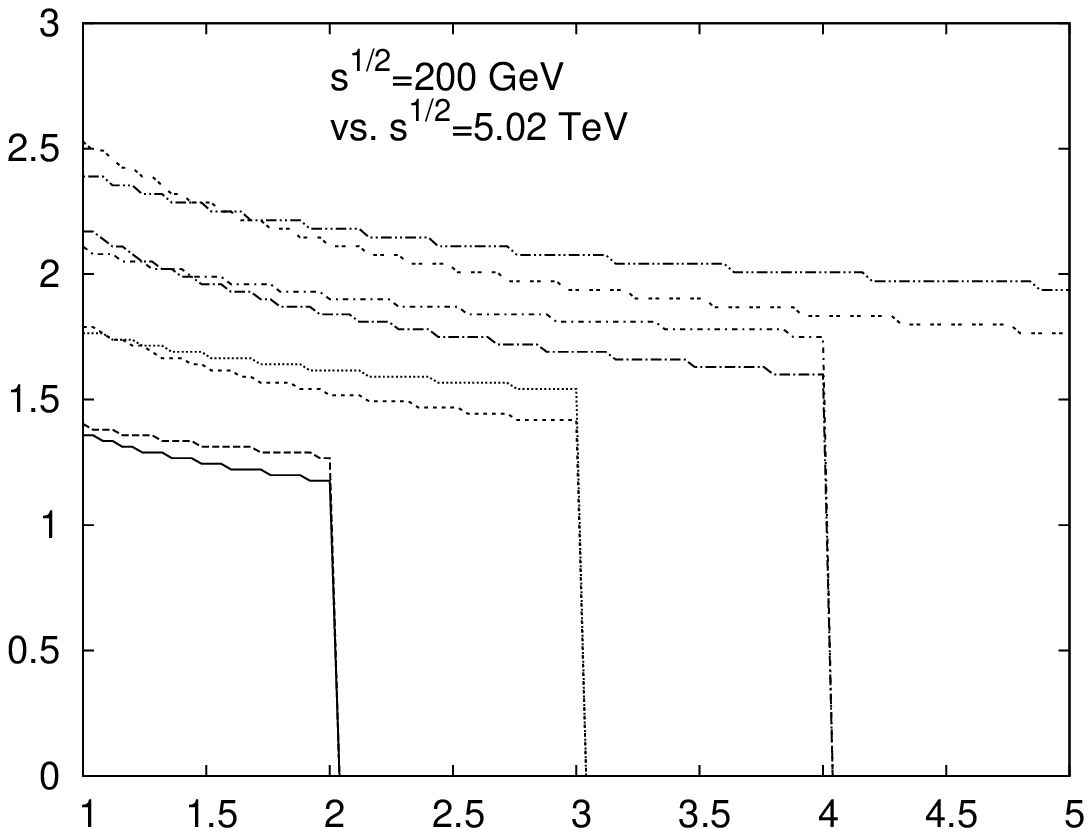, width=7 cm}
\caption{Powers $\alpha_n(x)$ at $\sqrt{s}=5.02$ TeV and
comparison of   $\sqrt{s}=200$ GeV and  $\sqrt{s}=5.02$ TeV.
Curves from bottom to top correspond to $n=2,3,4,5$. In the rifht panel
for each $n$ the upper curve corresponds to the higher energy.}
\end{center}
\label{fig4}
\end{figure}

Once these characteristics of cumulative strings formation are known one can start the
Monte-Carlo procedure to finally find the string distributions and cumulative
multiplicities. We performed 10000 runs in our Monte-Carlo program.
For pA collisions we choose p-Ta  at 27.5 GeV. This case at comparatively low energies is hardly suitable for our color string picture
(the length of cumulative strings is then restricted by $Y/2\sim 3$)
We choose it having in mind the existing old experimental data
on cumulative pion production ~\cite{bayu,niki}.  For AA we considered
Cu-Cu and Au-Au  collisions at 200 GeV (RHIC) and Pb-Pb collisions at 5.02 TeV (LHC).

\subsection {p-Ta at 27.5 GeV}
\begin{figure}
\begin{center}
\epsfig{file=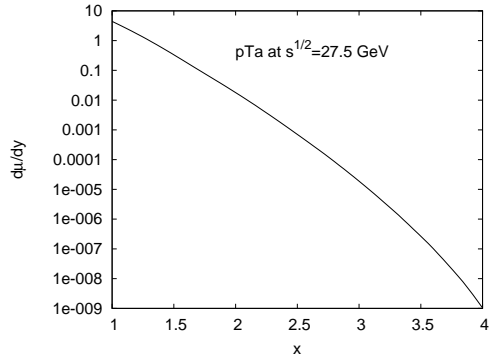,width=7 cm}
\caption{Multiplicities per unit rapidity for production of cumulative pions at cumulativity $x\geq 1$
in p-Ta collisions at 27.5 GeV.}
\end{center}
\label{fig6}
\end{figure}

The described numerical calculation gave the numbers of fused strings show (NFS) in Table.1
The data for $n=1$ give the number of non-fused strings and so with $x_1\leq 1$. It has been given only for comparison of
fused and non-fused strings. We repeat that the data refer only to the cumulative situation when the number of spring is restricted to two for each
nucleon in the nucleus. Should one leave this restriction the numbers will considerably grow and strongly increase with energy.
\vspace* {0.2 cm}
\begin{center}
{\bf Table 1}\\
\vspace* {0.2 cm}
\begin{tabular}{|r|r|}
\hline
$n$& NFS\\\hline
1&132\\\hline
2&57\\
3&21\\
4&6.3\\
5&1.6\\\hline
\end{tabular}
\end{center}
From these set of strings one obtains the multiplicities per unit rapidity show in Fig. \ref{fig6}

\subsection {AA}
In this case the distribution of cumulative strings and multiplicities depends on the impact parameter $b$.
We split roughly our results into three categories depending on the value of $b$:
central with $ 0\leq b\leq 0.4 R_A$, mid-central with $ 0.4 R_A\leq b\leq 0.8 R_A$
and peripheral with $b\geq 0.8 R_A$ where $R_A =A^{1/3} 1.2$ fm is the effective "nucleus radius".
The distribution of cumulative strings is shown in Tables 2.3 and 4 for collisions Cu-Cu and AuAu at 200 GeV and
Pb-Pb at 5.02 TeV.\\
\vspace* {0.2 cm}
\begin{center}
{\bf Table 2} Cu-Cu at 200 GeV\\
\vspace* {0.2 cm}
\begin{tabular}{|r|r|r|r|}
\hline
$n$& NFS central&NFS mid-central& NFS peripheral\\
\hline
1&166&36&22\\
\hline
2&20&7.5&3.2\\
3&4.8&1.5&0.40\\
4&0.96&0.2&0.042\\
5&0.16&0.039&0.030\\\hline
\end{tabular}
\end{center}
\vspace*{0.2 cm}
\begin{center}
{\bf Table 3} Au-Au at 200 GeV\\
\vspace* {0.2 cm}
\begin{tabular}{|r|r|r|r|}
\hline
$n$& NFS central&NFS mid-central&NFS peripheral\\
\hline
1&139&76&46\\
\hline
2&62&26&11\\
3&24&8.0&2.6\\
4&7.4&2.0&0.50\\
5&2.06&0.48&0.077\\\hline
\end{tabular}
\end{center}
\vspace*{0.2 cm}
\begin{center}
{\bf Table 4} Pb-Pb at 5.02 TeV\\
\vspace* {0.2 cm}
\begin{tabular}{|r|r|r|r|}
\hline
$n$& NFS central&NFS mid-central&NFS peripheral\\
\hline
1&148&80&47\\
\hline
2&66&27&12\\
3&25&8.5&2.8\\
4&8.2&2.3&0.52\\
5&2.26&0.47&0.079\\\hline
\end{tabular}
\end{center}

The corresponding multiplicities per unit rapidity are shown in Figs. 7 and 8.
\begin{figure}
\begin{center}
\epsfig{file=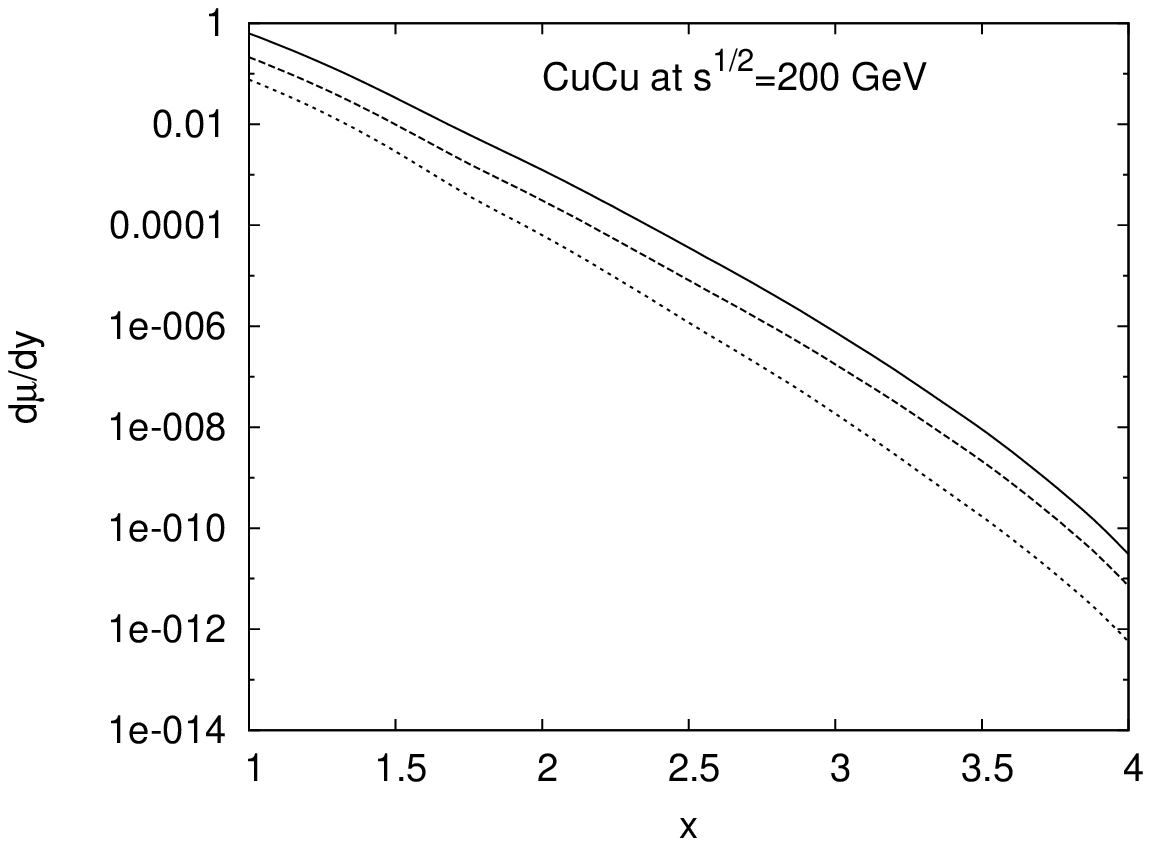,width=7 cm}
\epsfig{file=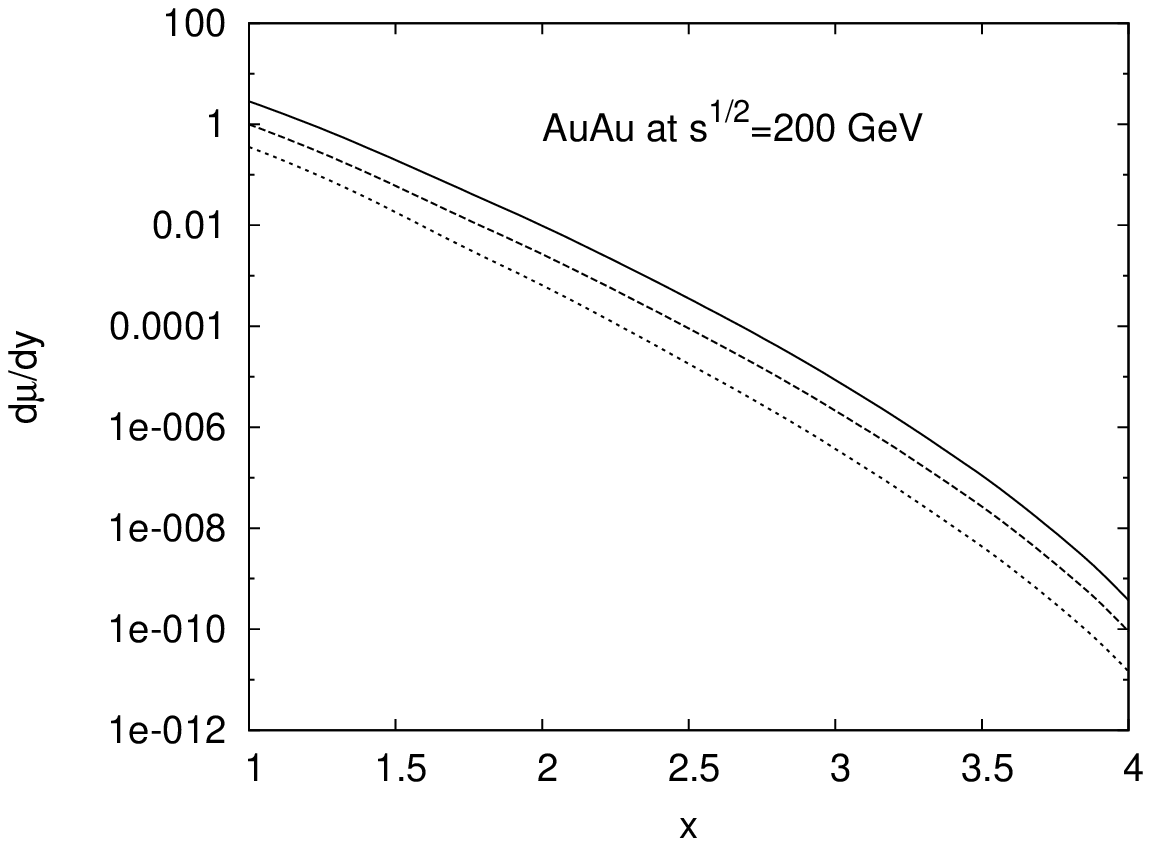,width=7 cm}
\caption{Multiplicities per unit rapidity for production of cumulative pions at cumulativity $x\geq 1$
in central (upper curve), mid-central (middle curve) and peripheral (lower curve) regions
in Cu-Cu and Au-Au collisions at 200 GeV.}
\end{center}
\label{fig7}
\end{figure}

\begin{figure}
\begin{center}
\epsfig{file=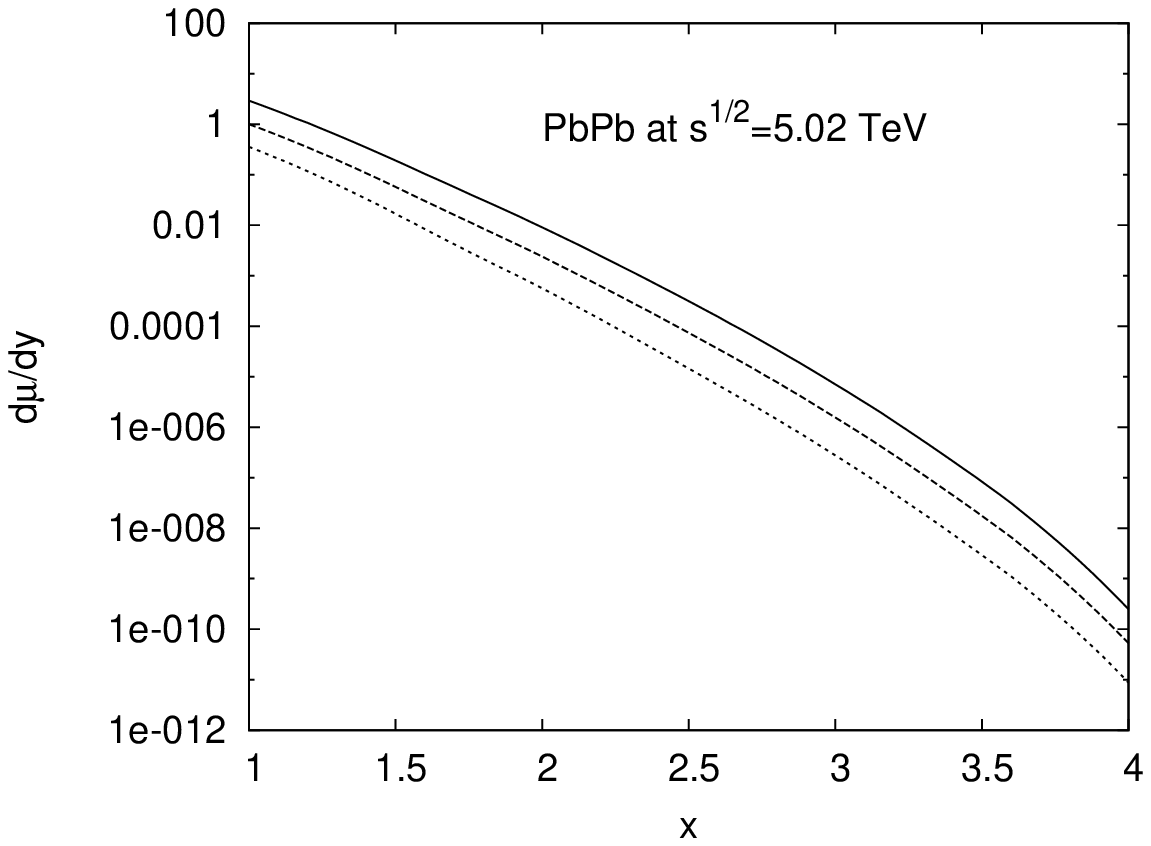,width=7 cm}
\caption{Multipliciies per unit rapidity for production of cumulative pions at cumulativty $x\geq 1$
in central (upper curve), mid-central (middle curve) and peripheric (lower curve) regions
in Pd-Pb collisions at 5.02 TeV.}
\end{center}
\label{fig8}
\end{figure}

The total multiplicities obtained after integration over all $b$ are illustrated in Figs. 9 and 10
\begin{figure}
\begin{center}
\epsfig{file=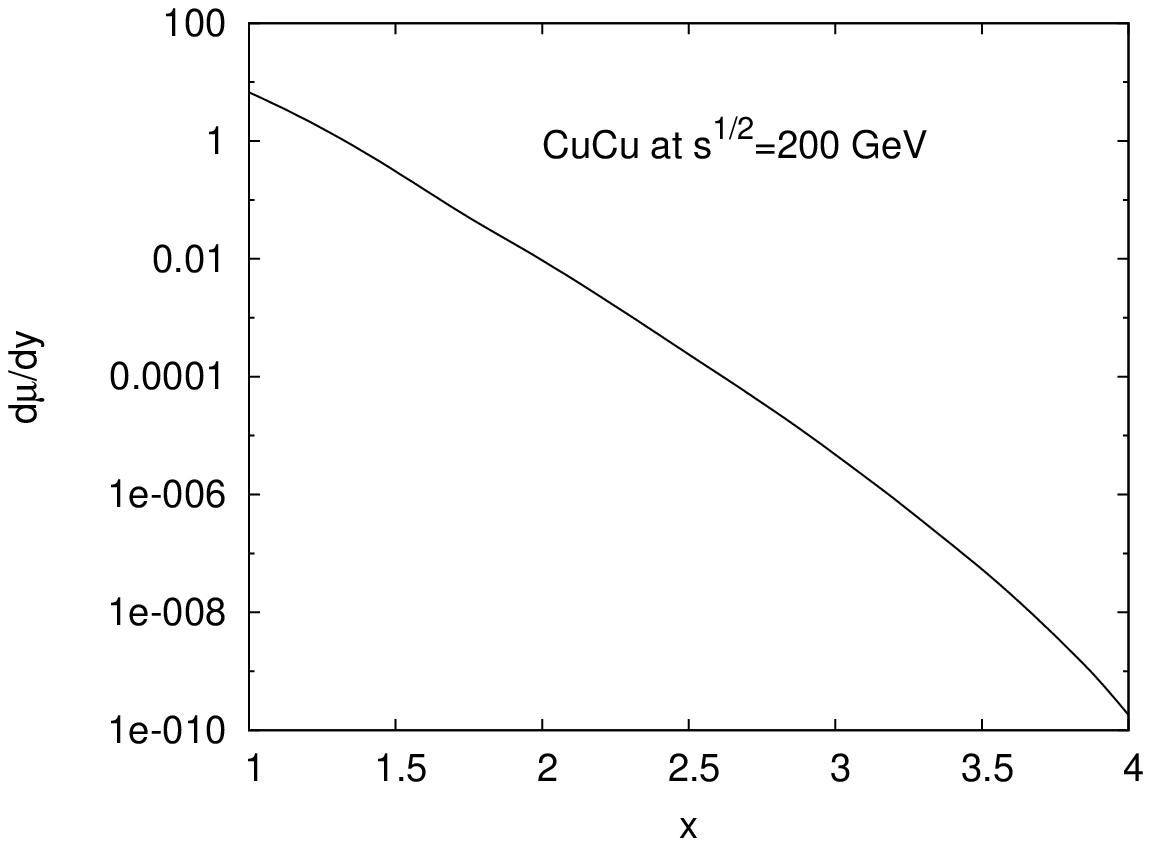,width=7 cm}
\epsfig{file=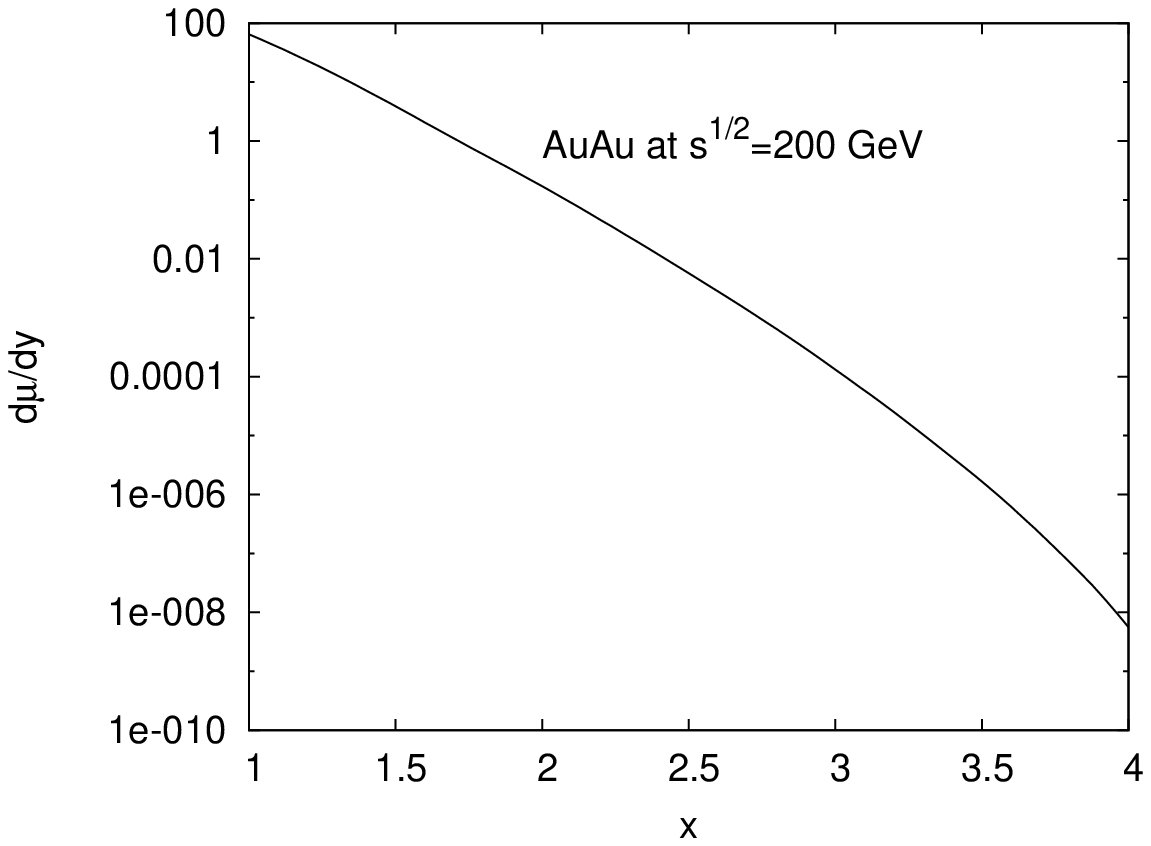,width=7 cm}
\caption{Total multiplicities per unit rapidity for production of cumulative pions at cumulativty $x\geq 1$
in Cu-Cu and Au-Au-Au collisions at 200 GeV.}
\end{center}
\label{fig9}
\end{figure}

\begin{figure}
\begin{center}
\epsfig{file=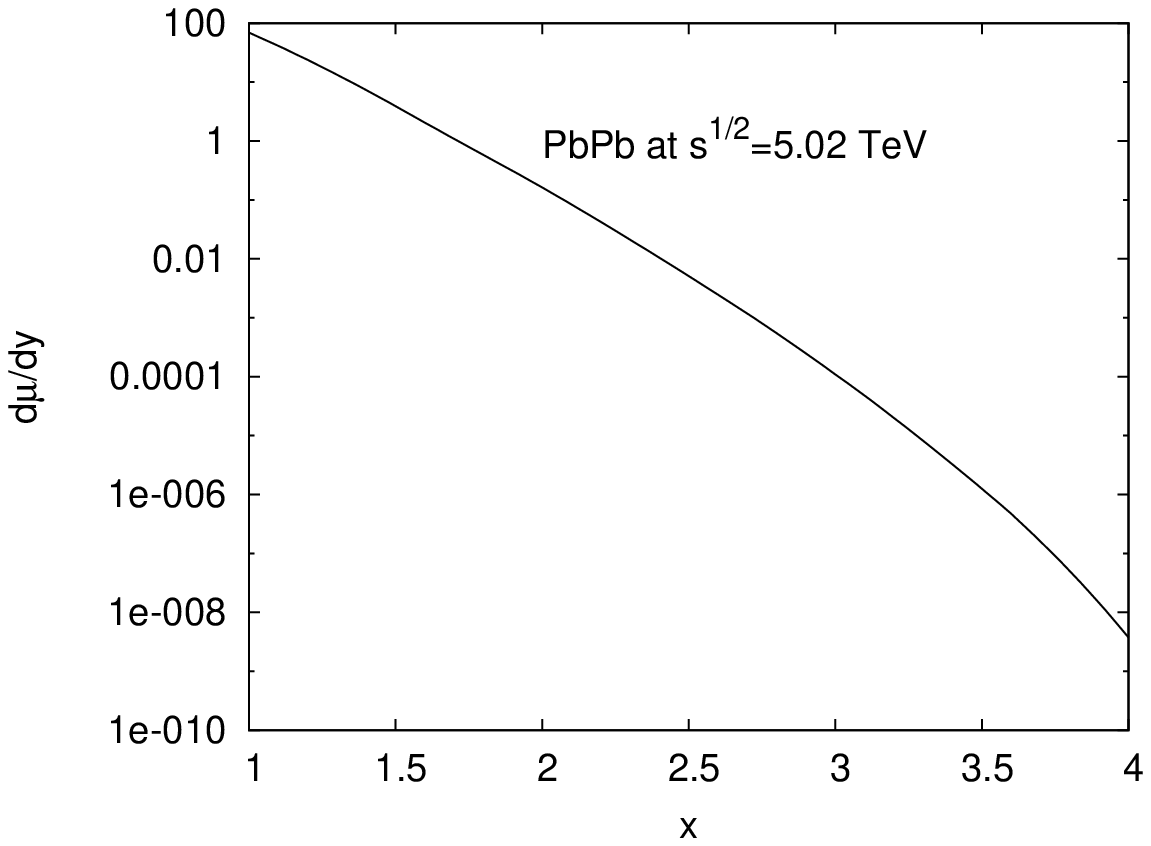,width=7 cm}
\caption{Total multipliciies per unit rapidity for production of cumulative pions at cumulativty $x\geq 1$
in Pb-Pb collisions at 5.02 TeV.}
\end{center}
\label{fig10}
\end{figure}

\section{Discussion}
In all cases our obtained multiplicities as a function of cumulativity have a simple exponential form at $1<x<3$
\beq
\mu(x)=C\exp^{-\beta x},\ \ 1<x<3.
\label{mux}
\eeq
The value of $\beta$ turns out to be of the order 5 and weakly varies for different cases.
For p-Ta at 27.5 GeV we find $\beta=5.0$. For Cu-Cu and Au-Au at 200 GeV  we obtain $\beta=5.6$ and 5.2 respectively.
Finally for Pb-Pb at 5.02 TeV we get $\beta=5.3$. We do not see any conclusive explanation for this small variation,
which may arise  from the difference in energy, nuclear wave function and insufficient number of runs.
As to the coefficient $C$ its values for p-Ta is 657 and for central  Cu-Cu, Au-Au and Pb-Pb are 166, 517 and 572 respectively.

Inspecting all cases we see that cumulative production of pions shows a large degree of universality, which is typical for the
fragmentation region of particle production.
The slope $\beta$ to a very large degree is explained by  overlapping of individual nucleons in the nucleus and roughly comes from the
probability to find $n>2$ nucleons occupying the same area in the nucleus. It is essential however that in the color string picture
overlapping of nucleons only occurs due to formation of strings and so interaction with the target. So comparing  to the initial very old
(not to say ancient) idea of the existence of "fluctons" in the nucleus with a larger mass and so capable to produce particles
in the cumulative kinematics the string picture does not see fluctons in the nucleus from the start.

Very recently experimental study of cumulative production was performed with Cu-Au collisions at 200 GeV ~\cite{bland}.
Cumulative jets were detected with the cm. energy $E$ greater than allowed by the proton-proton kinematcs $E<100$ GeV.
The data have been presented in two forms: raw data coming from the detectors and the so-called unfolded data which presumably take into account
distortions due to different sources of the concrete experimental setup. Remarkably in the cumulative region the raw data fall exponentially
as given by (\ref{mux}) with the slope $\beta\simeq 5.1$ which  does not practically depend on centrality nor jet characterisics.
Our results in this paper refer rather to collisions of  identical nuclei, such as Cu-Cu or Au-Au. But as we argued the cumulative production
in the projectile region does not depend on the target whose influence is only felt via the overlap in the transverse space. With different nuclei the
number of active nucleons will be different but in our picture it will influence only  the magnitude of the production rate but not
its $x$-dependence. The observed slope $\beta=5.1$  therefore agrees well with our predictions.

On the other side  the unfolded data have a different $x$ dependence in the power-form
\[\frac{dN}{N dE}=\Big(-\frac{E}{E_0}\Big)^p\Big(\frac{E}{E_0}\Big)^{-q}\]
with $p$ and $q$ adjusted to the data and $E_0=163$ or 193 GeV depending on the cone width of the jet but not on centrality.
These unfolded data do not behave in accord with  our predictions. This discrepancy may proceed from our simplfied picture of parton fragmentauion
(our partons go into pions with 100\% probability) and certainly deserves better study both of our treatment and experimental subtleties.

As mentioned in the Introduction cumulative pions were also seen in the HIJING and DPMJET models of Cu-Au collisions. Since the authors were
mostly interested in the overall spectra no special attention and analysis were
given to the cumulative region. However one can note that they give different predictions for cumulative pions (HIJING more paricles and with
greater energies) and that in HIJING rather strong centrality dependence was found. This latter property contradicts both our model
and experimental findings.

Note in conclusion  that  the flucton idea for cumulative production cannot be discarded altogether. One can envisage formation of a very fast particle already before the interaction.
As mentioned in the Introduction , cumulative production within this picture corresponds to the so-called spectator mechanism.
One expects in this approach that the leading particle will be one of the nucleons from the fast nucleus itself. The possibility of its
formation was
discussed in our old paper in the framework of a simple quark-parton model ~\cite{brave1}. It was later shown that for the cumulative protons
this spectator mechanism gives the bulk of the contribution and the direct mechanism considered in this paper  is suppressed ~\cite{brave2}.
This may explain why applied to the cumulative proton production in p-Ta collisions at 27.5 GeV the color string approach gave multiplicities
two orders of magnitude smaller  the experiment ~\cite{bfmp}.

\end{document}